# Numerical calculation of the lowest eigenmodes of the Laplacian in compact orientable 3-dimensional hyperbolic spaces.

J.P. Pansart[*]
Commissariat à l'Energie Atomique, CEN Saclay, DSM/IRFU/SPP,
91191  Gif-Sur-Yvette, France

Abstract.
A simple method to compute numerically the lowest eigenmodes of the Laplacian in compact orientable hyperbolic spaces of dimension 3 is presented. It is applied to the Thurston manifold, the Weber-Seifert manifold, and to the spaces whose fundamental domain is a regular icosahedron.

## 1  Introduction.

Compact hyperbolic spaces are chaotic system for classical trajectories, but as quantum systems, there exists well defined solutions. Although the Cosmic Microwave Background anisotropy data suggest a locally flat universe, nothing is really known about its global structure. Compact hyperbolic spaces may still be an attractive possibility, and there is an infinity of them. We refer the reader to the introduction of [1] for a complete discussion. The calculation of the observed anisotropies requires to know the eigenmodes of the Laplacian. In this note we present a simple method to compute numerically the lowest eigenvalues of the Laplacian (for scalar functions) in compact orientable hyperbolic spaces of dimension 3. The calculation is based on a $\chi^2$ method, as in [2], but uses a more geometrical approach. It also provides approximate eigenfunctions as expansions on known functions. The method is applied to the Thurston manifold, in order to compare with the results of [3], the Weber-Seifert manifold (which has a large volume), and to the spaces whose fundamental domain is a regular icosahedron. This last example allows to compare spaces having the same fundamental domain but completely different structures. Reference [3] uses the elegant "boundary element method", however the calculated eigenfunctions are purely numerical, and not presented as expansion on known functions.

## 2  Compact hyperbolic manifolds.

The hyperbolic n dimensional space $H^n$ is defined as the « upper part » of the sphere of radius $\sqrt{|K|}$ in the Minkowski space $M^{n+1}$. More precisely, if $\{x^\alpha\}$ are cartesian coordinates in $M^{n+1}$ with origine $O_M$, $H^n$ is the surface defined by :

---

[*] Retired from DSM/DAPNIA/SPP  since October 1, 2007. This work was begun at the end of  2006. Contact: jean-pierre.pansart@orange.fr

$$\sum_{\alpha=0}^{n-1} x^\alpha x^\alpha - x^n x^n = K$$

with : $K < 0$ and $x^n \geq \sqrt{|K|}$ . We set : $R = \sqrt{|K|}$ and use « spherical » coordinates. For $H^3$ in $M^4$ :

$$\begin{aligned}
x^0 &= R\, sh\chi\, c & ; \quad c = \cos(\theta)\,,\ s = \sin(\theta) \\
x^1 &= R\, sh\chi\, s\, c_\varphi & ; \quad c_\varphi = \cos(\varphi)\,,\ s_\varphi = \sin(\varphi) \\
x^2 &= R\, sh\chi\, s\, s_\varphi & \\
x^3 &= R\, ch\chi & \\
\chi &\geq 0\,,\ \theta \in [0,\pi]\,,\ \varphi \in [0,2\pi] &
\end{aligned} \qquad (1)$$

Then the linear element of $H^3$ is :

$$ds^2 = (dx^0)^2 + (dx^1)^2 + (dx^2)^2 - (dx^3)^2 = R^2\left[d\chi^2 + sh^2\chi\,((d\theta)^2 + s^2(d\varphi)^2)\right] \quad (2)$$

The coordinates $(\chi,\theta,\varphi)$ are the Riemann normal (spherical) coordinates with origine at $\chi = 0$ , which corresponds to the point $(0,0,0,R)$ in $M^4$ .

In the following $g_{\alpha\beta}$ is the metric tensor and $g^{\alpha\beta}$ its inverse, whatever the coordinate system used. The curvature tensor is : $R_{\alpha\beta\gamma\delta} = -\frac{1}{R^2}(g_{\alpha\gamma}g_{\beta\delta} - g_{\alpha\delta}g_{\beta\gamma})$ , the Ricci tensor is :

$R_{\alpha\beta} = -\frac{2}{R^2} g_{\alpha\beta}$ and the scalar curvature : $R_H = -\frac{6}{R^2}$ . From now on, we set $R = 1$ .

$H^n$ is simply connected. From theorem 2.4.10 of [4] : a manifold $M^n$ of dimension $n$ and negative constant curvature is complete and connected if and only if it is isometric to a quotient $H^n / \Gamma$ where $\Gamma$ is a group of isometries of $H^n$ acting freely and properly discontinuously.

$\Gamma$ is a subgroup of the component of $O(n,1)$ which preserves $H^n$ . It is called the group of deck transformation. $H^n$ is the universal covering of $M^n$ .

In the following we shall consider 3-dimensional compact and orientable manifolds only. The elements of the group $\Gamma$ are screw motions. A screw motion is the product of a transvection and a rotation. $H^n$ , which is of constant curvature, is a symetric space. A transvection in a symetric space is an isometry which generalises the notion of translation. It is defined as the product of two successive symetries whith respect to two different points $A$ and $B$ . The geodesic going through these two points is invariant and is called base geodesic. In $H^3$ one can perform a rotation around this geodesic. The transvection and this rotation commute and the base geodesic is invariant. The product of two screw motions is still a screw motion. In the following we shall call $\gamma^i$ the generators of $\Gamma$ and $g^i$ the corresponding base geodesics. The base geodesic of a motion $\gamma \in \Gamma$ is given by the intersection of the invariant plane, associated to the real eigenvalues of the matrix representing $\gamma$ in $M^4$ , with $H^3$ .

If we call $L$ the « length » of the transvection, which is twice the distance between $A$ and $B$, a point whose spherical coordinates are ($\chi, \theta, \varphi$, $c = \cos(\theta)$) is transformed, by a transvection along $Oz$, into a point of coordinates ($\chi', \theta', \varphi'$, $c' = \cos(\theta')$) given by :

$$ch(\chi') = ch(\chi)ch(L) + sh(\chi)sh(L)c$$
$$c' = (ch(\chi)sh(L) + c\,sh(\chi)ch(L))/sh(\chi') \qquad (3)$$
$$\varphi' = \varphi$$

We shall also use cylindrical coordinates wich we define as follows :
The $z$ coordinate of a point $M$ is given by the orthogonal projection of that point on the $Oz$ axis. The distance between $M$ and that projection is the radius $\rho$. The azimuth $\varphi$ is the same as for the spherical coordinates. The linear element is given by :

$$ds^2 = (d\rho)^2 + ch^2\rho\,(dz)^2 + sh^2\rho\,(d\varphi)^2 \qquad (4)$$

The two sets of coordinates are related by :
$$ch\chi = ch\rho\,chz \quad , \quad thz = c\,th\chi \quad , \quad sh\rho = s\,sh\chi$$

In cylindrical coordinates the action of a transvection along $Oz$ is simple and given by :
$$\rho' = \rho \quad , \quad z' = z + L \quad , \quad \varphi' = \varphi \qquad (5)$$

In the following, we shall call respectively $L_i$ and $\omega_i$ the length and the rotation angle of the screw motion $\gamma^i$.

The actions of infinitesimal transvections, whose base geodesics are the coordinate axis, and infinitesimal rotations around these axis are given in Appendix A for both coordinate systems.

## 3 Eigenmodes of the Laplacian.

We consider only scalar functions (real or complex), not differential forms, and call $\Phi$ the functions defined on $M^3$. The Lapacian is :

$$\Delta\Phi = \frac{1}{\sqrt{g}}\partial_\alpha(g^{\alpha\beta}\sqrt{g}\,\partial_\beta\Phi) \qquad (6)$$

where $g$ is the determinant of $g_{\alpha\beta}$.

The function $\Phi$ can be considered as a periodic function defined on $H^3$ (we give the same name) satisfying :

$$\Phi(\gamma\,x) = \Phi(x) \qquad (7)$$

for any point $x$ of $H^3$ and any element $\gamma$ of $\Gamma$. This condition is equivalent to :
$\Phi(\gamma^i\,x) = \Phi(x)$ for any generator $\gamma^i$ and their inverse. In particular, $\Phi$ must be periodic on the base geodesics.

Since the Laplacian commutes with any isometry we can search for functions which are eigenfunctions of the Laplacian and of some commuting subset of $\Gamma$. We shall consider commuting subsets of $\Gamma$ of the form $(\gamma^i)^p$ for some $i$ and $p \in \mathbb{Z}$. We shall choose, for $i$, the isometry which has the longest transvection, although this is a priori not necessary, it will be justified at the end of this section. We shall consider its base geodesic as the polar axis for spherical or cylindrical coordinates. For the rest of this note, the generators are renamed such that $\gamma^0$ becomes the one with the longest transvection and such that its base geodesic $g^0$ defines the *Oz* axis of the spherical or cylindrical coordinates (this is always possible by performing a transvection and a rotation around the origin).

In cylindrical coordinates functions of the form :
$$\Phi \sim I(\mu_z, \nu, \rho) \exp(i(\mu_z z + \nu \varphi)) \quad , \quad \nu \text{ integer} \quad (8)$$
are invariant under the action of $\gamma^0$ provided that : $\mu_z L_0 + \nu \omega_0 = 2\pi m_z$, where $m_z$ is an integer. This last constraint defines $\mu_z$.

At this stage one has two possibilities. Either one expands $\Phi$ on the known basis of eigenfunctions of the Laplacian in spherical coordinates (as in [2] for instance), or one keeps on using cylindrical coordinates.

*Spherical coordinates.*

With equation (2) in mind we naturally try to use a basis of eigenfunctions of $\Delta$ in spherical coordinates. The solutions are known and of the form ($\beta$ real) :
$$\Delta \Phi = -(1+\beta^2)\Phi \quad , \quad \Phi \simeq \phi_\beta^l(\chi) Y_l^m(\theta, \varphi) \quad (9)$$
where $Y_l^m(\theta, \varphi)$ are the usual spherical harmonic functions, $\phi_\beta^l \simeq \frac{1}{\sqrt{sh\chi}} B_\lambda^\mu(\chi)$, and $B_\lambda^\mu(\chi)$ are Legendre functions with $\lambda = -\frac{1}{2} + i\beta$ and $\mu = -\frac{1}{2} - l$ (see [5] for instance). The solution $\Phi = Constant$ corresponds to the 0 eigenvalue, and will not be considered anymore. These functions are regular and for example : $\phi_\beta^0 = \frac{\sin(\beta \chi)}{\beta \, sh\chi}$.

An infinitesimal transvection along *Oz* has the form (Appendix A) :
$$\Phi(x') = \Phi(x) + \varepsilon \, T_z \, \Phi \quad \text{where :} \quad T_z = c\partial_\chi + \frac{ch\chi}{sh\chi} s^2 \partial_c \quad (10)$$
The expression of the transvection component of $\gamma^0$ is the exponentiation of the operator $T_z$. A rotation by an angle $\omega$ around the *Oz* axis is simply : $\chi \to \chi$ , $\theta \to \theta$ , $\varphi \to \varphi + \omega$.

Using the recurrence relations for the Legendre functions [5], one obtains (Appendix C) :
$$(2l+1) T_z(\phi_\beta^l Y_l^m) = (l+m)\sqrt{\frac{2l+1}{2l-1}\frac{l-m}{l+m}} \phi_\beta^{l-1} Y_{l-1}^m$$
$$- (l+1-m)(\beta^2+(l+1)^2)\sqrt{\frac{2l+1}{2l+3}\frac{l+1+m}{l+1-m}} \phi_\beta^{l+1} Y_{l+1}^m \quad (11)$$

It is then possible to build functions that are eigenfunctions of $\Delta$ and $\gamma^0$. We write such functions as : $\Phi \simeq I(\mu_z, \nu, \rho) \exp(i(\mu_z z + \nu \varphi)) = \sum_{l,m} a_{lm} \phi_\beta^l(\chi) Y_l^m(\theta, \varphi)$. The orthogonality of the $e^{i\nu\varphi}$ on $[0, 2\pi]$ implies $m = \nu$, and by applying twice the operator $T_z$ to both members of the latter equation one has : $-\mu_z^2 \Phi = \sum_{l,m} a_{lm} T_z^2\left(\phi_\beta^l(\chi) Y_l^m(\theta,\varphi)\right)$ which, using (11), gives a matrix equation for the calculation of the $a_{lm}$ coefficients.

Unfortunately, although these equations are simple, their use in numerical calculations meets important problems as we explain now.
Let us consider the example of the Weber-Seifert manifold which has a high degree of symetry. Its fondamental domain is a regular dodecahedron. The transvection of the generators $\gamma^i$ have all the same length, and their rotation angles are all equal to $3\pi/5$. Their base geodesics go through the origin, and through the middle of the corresponding faces. The generators can be deduced from one another by simple rotation around the origin :
$\gamma^j = R \gamma^i R^{-1}$. Therefore it seems that spherical coordinates are well suited to study that case.
As said above, $\Phi$ must be periodic on the base geodesics of the generators $\gamma^i$, and, on them, we can expand $\Phi$ as : $\Phi = \sum r_j e^{i k_j \chi}$ where : $k_j = 2\pi j/L$ and $j$ is an integer. By analogy with the spherical Bessel functions of the Euclidean case, one can write :
$$e^{ik\chi} = \sum_l i^l (2l+1) b^l \, \phi_\beta^l(\chi) \tag{12}$$
By derivating twice both members of this equation and using the Legendre function recurrence relations, the $b^l$ cofficients are related by :
$$k(2l+1)b^l = (l+1)b^{l+1} + l(\beta^2 + l^2)b^{l-1} \tag{13}$$
$$b^0 = 1 \quad , \quad b^1 = k \quad , \quad b^1 = \tfrac{1}{2}(3k^2 - (1+\beta^2)) \quad , \quad ...$$
When one wants to implement this numerically, because of the shape of the functions $\phi_\beta^l(\chi)$, the number of terms necessary for $\chi$ larger than 2 increases very rapidly. But at high $l$ values and for $\chi$ smaller than or of the order of 1, the recurrence relations used to compute these functions fail, unless one uses 128 bits calculations. For these technical reasons we have preferred to work with cylindrical coordinates.
Note that, in the case $\nu = 0$, the coefficients $a_{lm}$ are simply :
$$a_{l0} \frac{Y_l^0(c=1)}{Y_0^0} = (2l+1)\, i^l \, b^l(k = \frac{2\pi m_z}{L})$$

*Cylindrical coordinates, radial functions.*

In cylindrical coordinates the Laplacian takes the form :
$$\frac{1}{sh\rho \, ch\rho} \partial_\rho(sh\rho \, ch\rho \, \partial_\rho \Phi) + \frac{1}{ch^2\rho} \partial_z^2 \Phi + \frac{1}{sh^2\rho} \partial_\varphi^2 \Phi = -(1+\beta^2)\Phi$$
and inserting (8) :
$$\frac{1}{sh\rho \, ch\rho} \partial_\rho(sh\rho \, ch\rho \, \partial_\rho I) - \frac{\mu_z^2}{ch^2\rho} I - \frac{\nu^2}{sh^2\rho} I = -(1+\beta^2)I \tag{14}$$

The radial function has the following properties :

For $\rho \to 0$ we set : $I = q(u)/ch\rho$ where : $u = sh\rho$. With that, (14) looks like a Bessel equation when $\rho \ll 1$, and one has the following expansion :

$$q = u^\alpha (1 + a_1 u + a_2 u^2 + ... + a_p u^p + ...) \tag{15}$$

with : $\alpha = \nu$ (in order to have a regular function at $\rho = 0$ ) , $a_{2p+1} = 0$ ,

$$4(\nu + 1) a_2 = \mu_z^2 + 1 - (\beta^2 + \nu^2) ,$$

$$8(\nu + 2) a_4 = a_2 (\mu_z^2 + 1 - \beta^2 - \nu^2) + (\beta^2 + \nu^2 - 2(\mu_z^2 + 1)) , ...$$

Equation (4) may suggest also to make the variable change $u = th\rho$ and try an expansion of the form :

$$I = u^\alpha (1 + a_1 u + a_2 u^2 + ... + a_p u^p + ...)$$

one obtains coefficients which are very similar to the above ones : $\alpha = \nu$ (in order to have a regular function at $\rho = 0$ ) , $a_{2p+1} = 0$ ,

$$4(\nu + 1) a_2 = \mu_z^2 + \nu^2 - (1 + \beta^2) ,$$

$$8(\nu + 2) a_4 = a_2 (\mu_z^2 + \nu^2 - (1 + \beta^2)) + (\mu_z^2 + \nu^2 - 2(1 + \beta^2)) , ...$$

but the former expansion (with $u = sh\rho$) has better convergence properties, in the parameter domain in which we have used them.

For $\rho \to \infty$ we set : $I = q(\rho)/\sqrt{sh\rho \; ch\rho}$ (this is suggested by the fact that, assymptotically, we expect an oscillating function and that $\sqrt{g} = sh\rho \; ch\rho$). The radial equation becomes :

$$\partial_{\rho\rho} q + \beta^2 q - \frac{(\mu_z^2 + \frac{1}{4})}{ch^2 \rho} q - \frac{(\nu^2 - \frac{1}{4})}{sh^2 \rho} q = 0 \tag{16}$$

We write : $q = A \cos(S)$ , the radial function has the following assymptotic expansion :

$$A = A_0 \left(1 + \frac{a_2}{ch^2 \rho} + \frac{a_4}{ch^4 \rho} + ...\right) \quad , \quad \frac{dS}{d\rho} = \frac{\beta}{(A/A_0)^2} \tag{17}$$

$$a_2 = \frac{\mu_z^2 + \nu^2}{4(1 + \beta^2)} \quad , \quad S = \varphi(\beta, \nu, \mu_z) + \beta \rho + 2 a_2 \beta (1 - th\rho) + ...$$

where $\varphi(\beta, \nu, \mu_z)$ is a phase , and $A_0$ is an overall normalisation constant which must be determined numerically.

The values of the radial function at low values of $\rho$ ( $\rho \le 0.25$) are calculated using the expansion (15) up to the twelvth order in $sh\rho$. For higher $\rho$ values, the differential equation (14) is solved numerically. Figure 1 shows some examples of radial functions and comparisons with the $\rho \to \infty$ assymptotic expansion up to terms of order $1/ch^8 \rho$ (included), where the phase and $A_0$ are numerically adjusted.

We could also have set : $I = q(u)/\sqrt{ch\rho}$ with $u = ch\rho$
In that case the radial equation looks like a Legendre equation when $\rho \to \infty$ :

$$(1 - u^2) \frac{d^2 q}{du^2} - 2u \frac{dq}{du} - (\beta^2 + \frac{1}{4}) q + (\mu_z^2 + \frac{1}{4}) \frac{1}{u^2} q + \frac{\nu^2}{u^2 - 1} q = 0$$

$$I \xrightarrow[\rho \to \infty]{} \frac{1}{\sqrt{ch\rho}} B^\mu_{-\frac{1}{2} + i\beta}(\rho) \quad , \quad \mu^2 = \mu_z^2 + \nu^2 + \frac{1}{4} \tag{18}$$

As far as numerical aspects are concerned, expansion (15) is the most convenient. Recurrences relations satisfied by the radial functions, similar to those of the Bessel functions, are given in Appendix B.

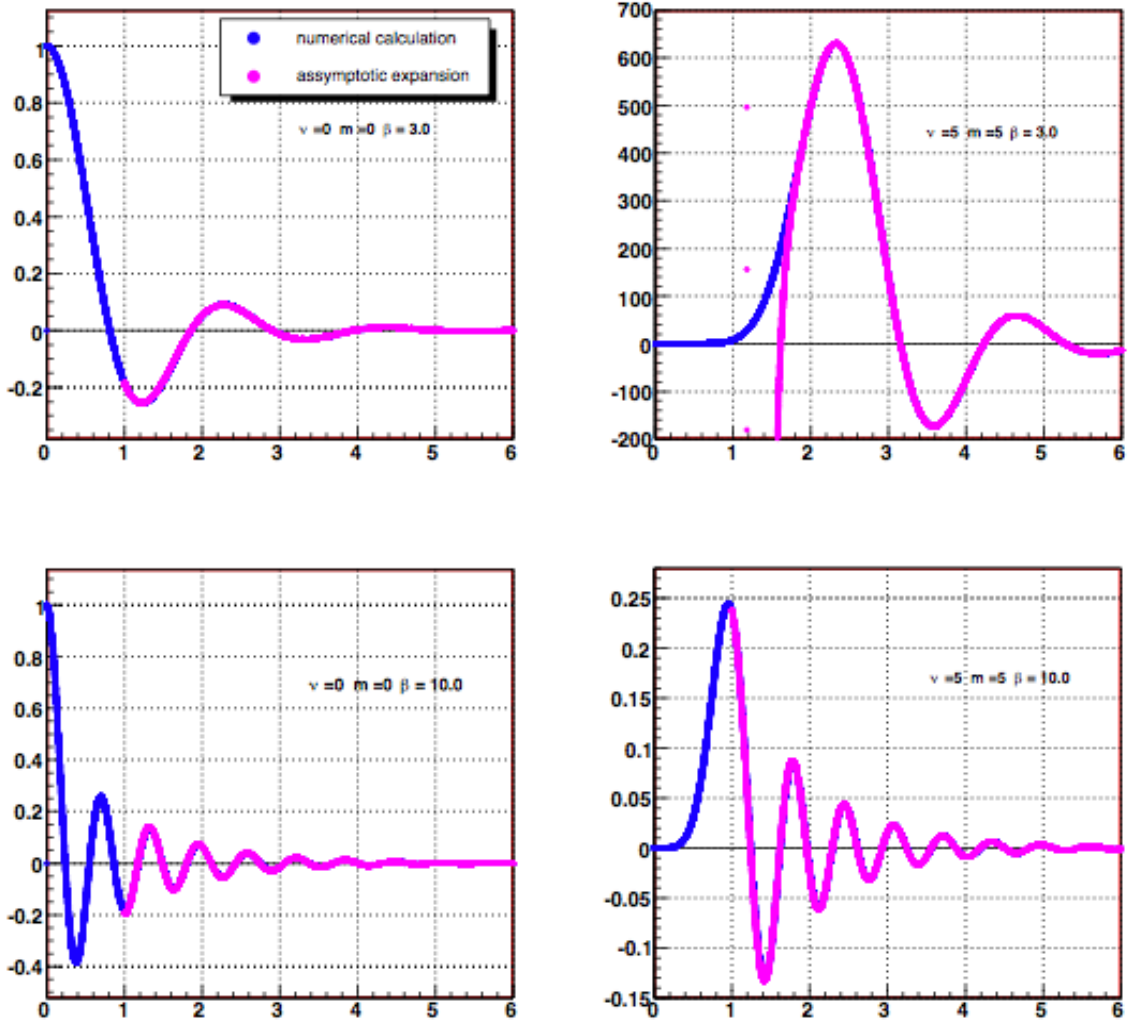

**Figure 1.** Examples of cylindrical functions.

We shall choose the following basis of real functions :
$$\Phi_{0,m,\nu} = I(\beta,\mu_z,\nu,\rho) \cos(\mu_z z + \nu\varphi) \tag{19a}$$
$$\Phi_{1,m,\nu} = I(\beta,\mu_z,\nu,\rho) \sin(\mu_z z + \nu\varphi) \tag{19b}$$
where : $\mu_z L_0 + \nu\omega_0 = 2\pi m_z$ , $m_z$ integer .

These functions are orthogonal on any cylinder of axis $Oz$ and height L (or a multiple). This is important for the least square method used to find solutions satisfying the periodicity relations (7), as described in section 5.

The action of the elementary motions on these cylindrical functions is given in Appendix C.

A first consequence of the assymptotic expansion (17) is that $\Phi$ can not be simply of the form (19). This can be seen by using again the example of the Weber-Seifert manifold. $\Phi$ must be periodic on the base geodesics of the generators $\gamma^i$, $i \neq 0$ which go through the origine $O$, that is to say, $\Phi$ must be a periodic function of $\chi$. When $\chi \to \infty$, $z \to z_{\lim}$ such that $thz_{\lim} = 1/\sqrt{5}$ and $\chi \to \rho + Ct$. The expansion (17) shows that functions of the form (19) can not satisfy the periodicity condition since they are decreasing when $\rho$ increases. Analogously to (12), $\Phi$ must be a sum :

$$\Phi = \sum_{m,\nu} c^{m,\nu} \phi_{m,\nu} \tag{20}$$

with $m \geq 0$, because if both $m$ and $\nu$ change sign, $\phi_{m,\nu}$ remains the same up to a sign. The sum (20) is also a consequence of (D3).

In practice, one can only calculate a limited number of terms in the sum (20), and the method used will be described in section 5. Since we consider only manifolds without border, we can use Rayleigh's theorem [6], which puts some constraints on the eigenvalues, by saying that the lowest eigenvalues are associated to the states having the lowest « kinetic energy ». In the remaining of this section we consider functions $f$ and $g$ defined on a manifold $M$ (without border, by hypothesis) which satisfy (using (6)) :

$$\int_M f^+ \Delta g \, dV = -\int_M \partial_\alpha f \, g^{\alpha\beta} \, \partial_\beta g \, dV = \int_M (\Delta f)^+ \, g \, dV \tag{21}$$

where the second integral is the scalar product of the gradients of the functions.

Rayleigh's theorem states that [6] :
Let $\lambda_1 \leq \lambda_2 \leq ...$ be the eigenvalues repeated the number of times equal to their multplicity. Then for any function $f \neq 0$ satisfying (21), we have : $\lambda_1 \leq (\int \nabla f . \nabla g \, dV)/\|f\|^2$ with equality if and only if $f$ is an eigenfunction of $\lambda_1$. If $\phi_1, \phi_2, ...$ is a complete orthonormal basis such that $\phi_j$ is an eigenfunction of $\lambda_j$, then for $f \neq 0$ satisfying : $(f, \phi_1) = ... = (f, \phi_{k-1}) = 0$ we have : $\lambda_k \leq (\int \nabla f . \nabla g \, dV)/\|f\|^2$, with equality if and only if $f$ is an eigenfunction of $\lambda_k$.

In order to use directly this theorem, one needs to construct functions on $H^3$ which satisfy the condition (7). One possibility woulb be, given any function $h$ on $H^3$, to write :
$f(x) \sim \sum_{\gamma \in \Gamma} h(\gamma x)$, but this would be very time consuming.
In fact we shall use this theorem only qualitatively. The inequalities say that the lowest eigenvalues will be best constrained if we use functions which do not vary rapidly, or, in other words, that the main contribution in (20) should come from the low $m$ and $\nu$ values of the basis functions (19). This justify the use, in practice, of a limited expansion in (20).
Note that, since the first eigenfunction is the constant function associated to the eigenvalue 0, if $\Phi$, in equation (20), is periodic on $H^3$, it is orthogonal to the first eigenfunction on $M$ by (21), and the requirements of Rayleigh's theorem are automatically satisfied for the second eigenfunction.
The Rayleigh's theorem justifies our choice of the base geodesics associated to the longest transvection as polar axis, since the solutions with the longest wavelengthes have the smallest "energy", at least in the vicinity of the base geodesic.

## 4 Description of the manifolds used as examples.

*Thurston manifold.*
$\Gamma$ has 8 generators, and the fundamental domain domain has 16 faces. It is the second smallest compact hyperbolic manifold known today, with a volume of $\sim 0.98$. (the smallest known manifold being the Weeks manifold with 9 generators and a volume of $\sim 0.94$).
The spectrum of the Laplacian eigenvalues has been calculated in [3] using the "boundary element method" and in [2] using a much more direct method. We shall compare our results with those of these references.

*Weber_Seifert manifold.*
It has been briefly described above. The fundamental domain is a regular dodecahedron. Its high degree of symetry, and the fact that the base geodesics of the generators all go through the origine makes it an attractive toy for understanding and tentative analytic calculations. The Weber-Seifert manifold is not the only one whose fundamental domain is a regular dodecahedron, there are 7 others (see [7]). In the tiling of $H^3$, each edge is common to 5 dodecahedrons and each summit is common to 20 volumes. The volume of the fondamental domain is $\sim 11.2$. The distance $h$ between the center of the fundamental domain and the center of the faces is : $ch^2(h) = (5+2\sqrt{5})/4$. There exists ideal dodecahedron (whose summits are at infinity but having finite volume), but this case will not be considered here.

*Manifolds whose fundamental domains is a regular icosahedron.*
They have 10 generators and the fundamental domain has 20 faces and 12 summits. There are 7 compact mainfolds. The tables of Appendix E show how the faces and their summits are identified. The volume is $\sim 4.69$. The distance $h$ between the center of the fundamental domain and the center of the faces is : $ch^2(h) = 3/2(3-\sqrt{5})$. There is no ideal icosahedron. In the tiling of $H^3$, each edge is common to 3 icosahedrons and each summit is common to 12 volumes.

## 5 Numerical calculation of approximate solutions.

For any value of $\beta$, the functions defined by equation (19) are eigenfunctions of both the Laplacian and $\gamma^0$ (wich is such that its based geodesic $g_0$ defines the $Oz$ axis), but the solutions must also satisfy all the constraints (7). In order to find approximate solutions, $\Phi$ is developped, as in (20), on the basis of cylindrical functions (19) using a limited number of terms, and the constraints (7) are enforced using a $\chi^2$ method to determine a set of coefficients $c^{m,v}$. Then we look at the behaviour of the function $\chi^2(\beta)$ to find satisfying solutions. The fact that the functions (19) forms an orthogonal basis, on cylinders of axis $Oz$, allows a good mode decoupling.
The calculations are done as follows :
- points $x_j$ are chosen randomly in a volume containing the fundamental domain, then their images $y_j^k = \gamma^k x_j$ are calculated for all the generators $\gamma^k$, $k \neq 0$ and their inverse.

- Since $\Phi$ is defined up to a multiplicative constant, we set either : $\Phi(O)=1$ , or, for some arbitrary point $A$ on $Oz$ : $\Phi(O)=0$ , $\Phi(A)=1$ .
- We define : $\chi^2 = \sum_i (\zeta_i - c^{m,v} \psi_{m,v})^2 / \sigma_i^2$ , where the index $i$ corresponds to a pair $(x_j, y_j^k)$. If one wants to enforce the periodicity condition one sets : $\zeta_i = 0$ and $\psi_{m,v} = \phi_{m,v}(y_i) - \phi_{m,v}(x_i)$ where the $\phi_{m,v}$ are the basis functions (19) . For the normalisation constraints : $\psi_{m,v} = \phi_{m,v}(x_i)$ and $\zeta_i = 0$ or 1 . $\sigma_i$ is an arbitrary error.

The calculations are local, which means that :
- both elements of the couple $(x_j, y_j^k)$ must be inside the fondamental domain or close to it.
- The periodicity conditions are imposed for the generators $\gamma^k$ (and their inverse) not for all the elements of $\Gamma$.

The reason for local calculations is due to the behaviour of the radial function $I(\beta, \mu_z, v, \rho)$. At small radii it is given by (15). The position of the first maximum increases if $v$ increases. In other words, the larger the working volume the larger the number of modes one has to consider.
The covariance matrix is inverted using the SVD decomposition of the SGL library.
In the following we have not tried to find the correct multiplicity of each eigenvalue. This could be done by solving directly the equations (7) for the above constraints using the SVD decomposition.

## 6 Results

For the following calculations, the parameter values used are typically $|v| \leq 20$ , $m \leq 5$ . In order to impose the periodicity conditions (7), points were chosen in a cylinder of radius 2, and their image should lie in a cylinder of radius ranging from 1.4 to 2. The number of these points is chosen to be $\sim 4000$. With such representative parameters, the calculation time necessary to explore the range $0 \leq \beta \leq 9$ by steps of 0.025, is around 140 mn on a MacBookPro, without special efforts to optimise the program.

*Thurston manifold.*
For the Thurston manifold, an example of the function $\chi^2(\beta)$ is shown in figure 2 for the case $\Phi(0)=1$ . Our results are compared with those of [3] in the following table. The first column reproduces Table 1 of [3], and the second column gives (-) the eigenvalues of the Laplacian, which is $-k^2$ in the notations of [3]. The third column gives the value of $\beta$ for which we have found a solution. If there is no correspondance, that means we have not found any convincing solution for this value of $k$ . The uncertainties on the values of $\beta$ are of the order of 0.01, and depends only on the $\beta$ step used to study the behaviour of $\chi^2(\beta)$ . The last column gives the type of solution found according to section 5. $\Phi(0)=0$ does not necessary means that the solution is antisymmetric with respect to the origin on the $Oz$ axis.

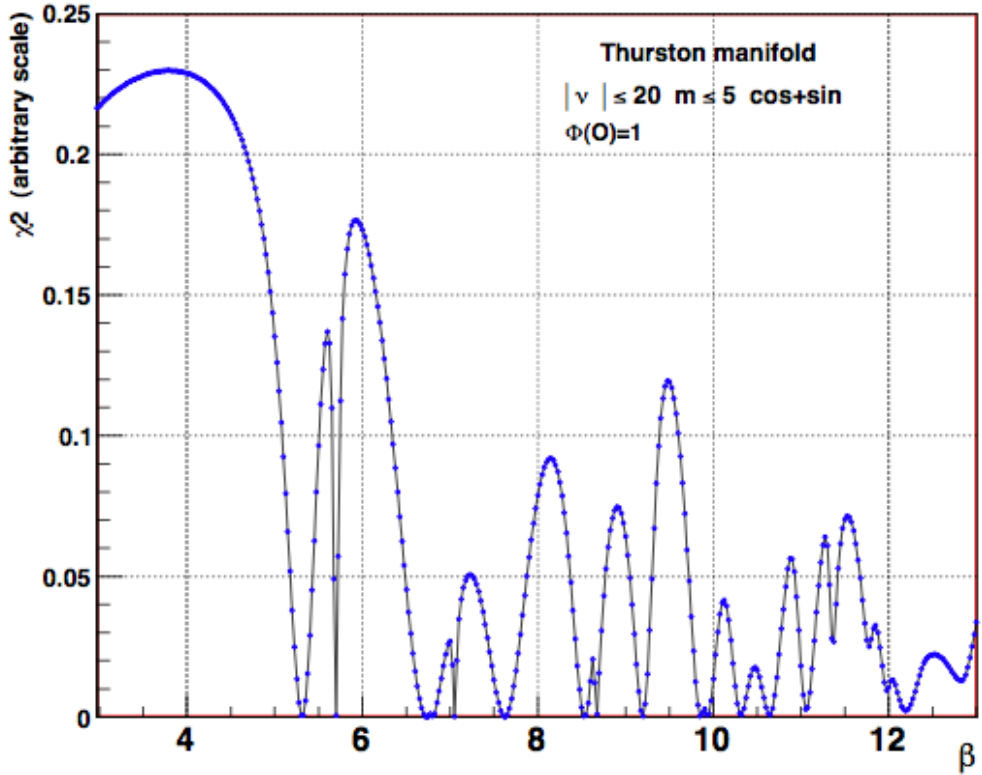

**Figure 2.**

| $k$ [3] | $k^2$ | $\beta$ | $1+\beta^2$ | $\Phi(0)$ |
|---|---|---|---|---|
| 5.41 | 29.27 | 5.31 | 29.2 | 1. |
| 5.79 | 33.5 | 5.70 | 33.5 | 1. |
| 6.81 | 46.4 | 6.72 | 46.2 | 1. |
| 6.89 | 47.5 | 6.80 | 47.2 | 1. |
| 7.12 | 50.7 | 7.06 | 50.8 | 1. |
| 7.69 | 59.1 | 7.62 | 59.1 | 0. , 1. |
| 8.30 | 68.9 | | | |
| 8.60 | 74. | 8.53 | 73.8 | 1. |
| 8.73 | 76.2 | 8.67 | 76.2 | 0. |
| 9.26 | 85.7 | 9.20 | 85.6 | 0. , 1. |
| 9.76 | 95.3 | | | |
| 9.91 | 98.2 | 9.85 | 98. | 1. |
| 9.99 | 99.8 | 9.94 | 99.8 | 1. |

We have not found solutions for the eigenvalues 68.9 and 95.3 while they are found both in [2] and [3]. We have used a smaller $\beta$ step, change the set of parameters (while keeping their number under 500 for practical reasons), even the base geodesic used as polar axis. In all the cases the eigenvalues found were perfectly stable, even if, sometimes, less visible. We can see the first of the missing eigenvalues but not the second one. We think that this due to the fact that we have limited the expansion of the solution on too small a set of functions.

The shape of the function for the lowest eigenvalue is rather jittery. For the lowest eigenvalue, $\Phi(z)$ on the $Oz$ axis is approximately constant, while for the two next ones, it oscillates with a period equal to the length of the transvection of $\gamma^0$ which is $L \simeq 1.04$.

*Weber-Seifert manifold.*

The function $\chi^2(\beta)$ is shown in figure 3 for the case $\Phi(0)=1$. Except for the lowest $\beta$ values, the structures of $\chi^2(\beta)$ are less visible than in the case of the Thurston manifold. This is due to the fact that the Weber-Seifert manifold has a large volume. In the expansion (20) the basis functions must be computed at larger radii, but $\phi_{mv}$ with larger $v$ becomes non negligible, and should be included in the calculations, which would increases the computation time.

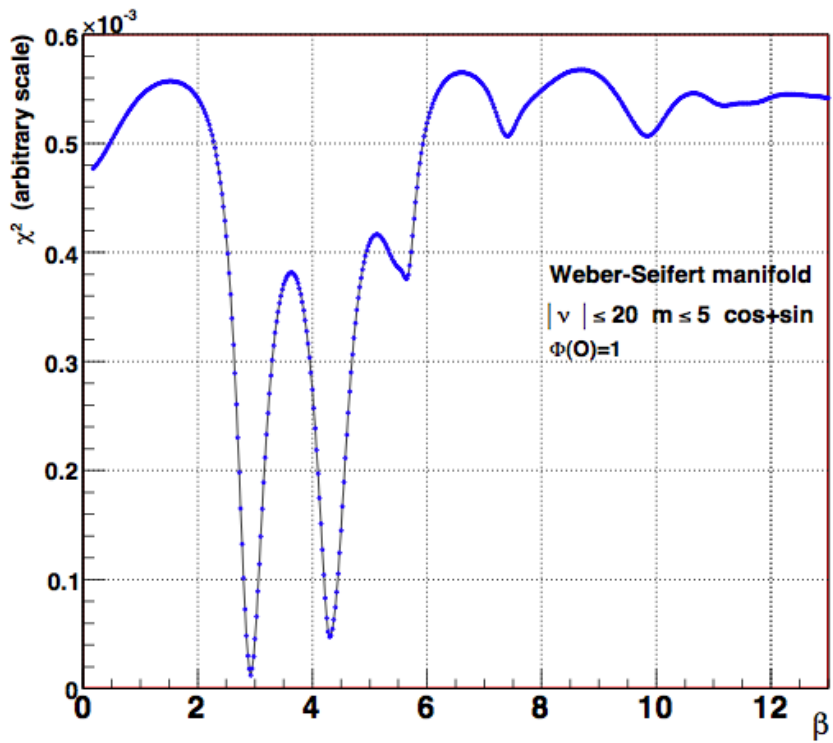

**Figure 3.**

The structures of the function $\chi^2(\beta)$ are not very clear, howewer, if we require the solutions to be symetric with respect to rotations around the $Oz$ axis with angles multiple of $2\pi/5$ (that we can impose by requiring $v$ to be a multiple of 5 in (20)), they all appear very neatly. In this case we have used $|v| \leq 30$, $m \leq 10$. The first eigenmodes are listed in the following table. The five first are such that the periodicity on the $Oz$ axis is $L$. On the $Oz$ axis, the first eigenfunction is well described by $\Phi \sim 0.89 + 0.12 * \cos(2\pi z/L) - 0.012 * \cos(4\pi z/L)$. The shape of this function on the other base geodesics is of the form : $\Phi \sim a + b\cos(2\pi\chi/L + phase)$ which shows that it is a « low energy » solution.

| $\beta$ | $\Phi(O)$ |
|---|---|
| 2.925 | 1 |
| 3.79 | 0 |
| 4.275 | 1 |
| 5.58 | 1 |
| 5.65 | 0 |
| 5.775 | 0 |
| 6.57 | 0 |
| 6.925 | 0 |
| 7.225 | 1 |

Although not detailed here, one can use the expansion (9) in the neighbourhood of the origin to show that there is no solution having the full symetry of the Weber_Seifert manifold. In fact, there is no solution with the same $\Phi(\chi)$ on all the base geodesics associated to the generators.

*The fundamental domain is an icosahedron.*
The results are presented in the following table, and an example is given in figure 4.
The first line is the number of the space described in appendix E. The second line gives the length of the longest transvection. Each column gives the $\beta$ values for which there is a solution. A question mark means that the situation is not clear, at least for the limited expansion used. The letter a means a solution of type $\Phi(0) = 1$, and the letter b $\Phi(0) = 0$. Here again a question mark means that the type is uncertain.

|   | 1 | 2 | 3 | 4 | 5 | 6 | 7 |
|---|---|---|---|---|---|---|---|
| $L$ | 1.629 | 1.465 | 1.629 | 1.736 | 1.008 | 1.736 | 1.736 |
| $\beta$ | 2.525 a | 2.675 a | 2.925 a | 3.40 b | 2.47 a | 3.40 b | 3.40 a |
|   | 3.275 a | 2.925 a | 3.27 a | 4.125 a | 4.33 a | 4.125 a | 3.525 a? |
|   | 4.11 a | 4.125 a | 4.05 b | 5.45 a | 5.50 a ? | 5.45 a | 4.125 a |
|   | 4.39 b | 4.48 a | 4.125 a | 5.75 b | 6.75 a | 5.75 b | 4.29 b |
|   | 4.60 a | 5.25 a | 4.25 b | 5.90 ? b |   | 5.925 ? b | 5.45 a |
|   | 5.075 a | 5.50 b? | 4.84 b | 6.60 ? b |   | 6.50 ? b | 5.75 a? |
|   | 5.43 a | 5.80 a? | 5.10 ? a | 7.02 b |   | 6.60 b |   |
|   | 5.60 b |   | 5.175 b | 7.82 b |   | 7.025 b |   |
|   |   |   | 5.45 a | 8.11 a |   | 7.79 b |   |
|   |   |   | 6.02 a |   |   | 8.10 a |   |

In all the cases, for the lowest eienvalues, the field $\Phi$, on the base geodesic used as coordinate axis, is well approximated by functions of the form : $a + b\cos(2\pi z/L + phase)$, where $a$ and $b$ are constants, and $L$ is the length of the transvection on this geodesic, which means that these functions have « low energy ».

There does not seem to be any correlation between the lowest values of $\beta$ and the length of the longest transvection, the whole structure of $\Gamma$ is involved although all these spaces have the same fundamental dommain. The structures of the functions $\chi^2(\beta)$ are not always as clear as in the case of the Thurston manifold. This is the case for spaces number 2, 3, 5 and 7. As already said, this is mainly due to the limitation in the expansion (20), which could be overcome by increasing the computation time.

We have repeated some of the calculations by changing the generator whose base geodesic defines the $Oz$ axis, even if it has a shorter transvection, with identical results.

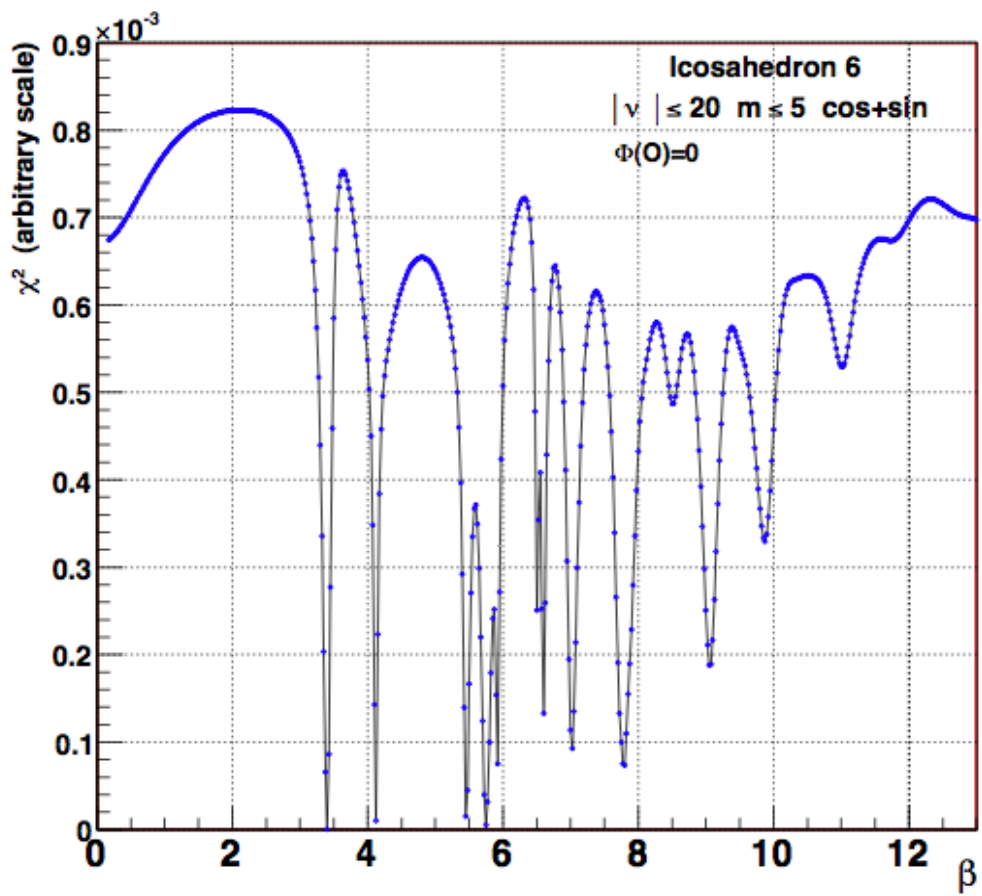

**Figure 4.**

## 7 Comments on the method and other attempts.

Whatever the method used to find the eigenvalues of the Laplacian of scalar functions on a compact manifold $M = H^3/\Gamma$, one has to solve the periodicity condition (7) on $H^3$.
In this section we explain why the most direct method, used in these notes, works for the lowest eigenvalues search. This method is not elegant. and although it is usual to present only what works and not what failed, we shall sketch other attempts which were unsuccessfull. The discussion uses only the Weber Seifert manifold for the reasons explained formerly in the text and in Appendix D. We focus our discussion on the determination of the first eigenvalue.

Using the corollary of the min-max theorem of [6] (p. 27) it is easy to find an upper bound for the first non zero eigenvalue. The spherical radial function $\phi_\beta^0$ (see section 3) is a solution of equation (9) for the Dirichlet problem on a sphere of radius $\pi/\beta$. Considering the largest sphere inscribed in the fundamental domain of the Weber-Seifert manifold, which has a radius equal to the distance between the centre of this domain and the center of the faces of the dodecahedron (section 4), the min-max theorem gives the following upper bound for the first non zero eigenvalue : $\beta \leq \pi/0.9964$.

In order to obtain a better estimate, we look for solutions of the form (20) : $\Phi = \sum_{m,\nu} c^{m,\nu} \phi_{m,\nu}$ where the sum is infinite. In practical calculations, the number of terms of this sum must be limited. One can use the basic symetry properties of the manifold in order to reduce the number of terms. One can check that the following sum :

$$\Phi = \sum_{m,\nu_1} c^{m,\nu_1} I(\beta,\mu_z,\nu_1,\rho) \cos(\mu_z z + \nu_1 \varphi) + \sum_{m,\nu_2} c^{m,\nu_2} I(\beta,\mu_z,\nu_2,\rho) \sin(\mu_z z + \nu_2 \varphi)$$

where $\nu_2$ is an odd multiple of 5 : $\nu_2 = (2p+1)5$, and $\nu_1$ a multiple of 10, satisfies the following conditions :
- the solution is invariant by rotations around the $Oz$ axis of angle $2\pi/5$.
- the function has the same value for almost all the summits of the fundamental dommain. There remains only one condition to satisfy, for instance (see the figure of Appendix E) : $\Phi(11) = \Phi(0)$.
- the function has the same value at the center of opposite faces, where each couple of faces corresponds to a generator.
- due to the generator properties, the edges of the dodecahedron can be grouped in six sets of five elements each, the elements of each set being in direct correspondance under the action of one of the element of $\Gamma$. If each edge is identified by its summit numbers, these sets are : (0-1, 16-19, 18-15, 9-3, 4-10), (0-3, 16-12, 8-14, 17-19, 6-2), (1-5, 19-18, 13-7, 2-0, 11-17), (5-8, 18-13, 12-6, 4-1, 14-15), (3-8, 12-13, 10-16, 5-11, 7-9), (4-6, 14-17, 9-15, 2-7, 11-10). The above expansion allows the function value to be the same at the edge middles of any given groups if the 2 following conditions are satisfied. For instance for the second group, that the function has the same value for the middle of the first and the fourth and fifth edges of the group.
The latter expansion is not the most general solution, because it requires that the above symetries are satisfied at each order, but one can check, a posteriori, that it provides a solution for the lowest eigenvalues. However it satisfies the basic symetries for each value of $\beta$.
In order to obtain a reasonable and unambiguous estimate of the first eigenvalue, one must use : $\nu = 0, \pm 5$, $m \leq 4$, that is to say 14 parameters. Imposing the condition : $\int_M \Phi \, dV = 0$, and fixing the value of the function at the origin reduces the number of parameters to 12.
In order to get a neat minimum of the $\chi^2$ we need to add at least 20 other constraints, and this gives $\beta \approx 2.97$. Not too bad an estimate for this low number of coefficients, but from a computational point of view, there is no real gain obtained by reducing the number of parameters down to the minimum possible.

As already said, the direct method used to solve condition (7) works but is not elegant. We now sketch briefly another attempt.

According to (D3) and (20), the condition (7) can be written :

$$\phi(\gamma_1 x) = c^{mv} \phi_{mv}(\gamma_1 x) = c^{mv} a_{mv}{}^{\sigma_1 v_1} \phi_{\sigma_1 v_1}(x) = \phi(x) = c^{mv} \phi_{mv}(x) \qquad (22)$$

and if we admit that the functions $\phi_{\sigma_1 v_1}$ are linearly independent, one gets the condition :

$$c^{mv}(a_{mv}{}^{\sigma_1 v_1} - \delta_m^{\sigma_1} \delta_v^{v_1}) = 0 \qquad (23)$$

where the notation $\delta_m^{\sigma_1}$ is certainly abusive but understandable. In order to implement this method one has firstly to compute the cofficients $a_{mv}{}^{\sigma_1 v_1}$. These coefficients are calculated by expanding the exponential (D2) and by using the relations of Appendix C. Unfortunateley, beacause of the large values of the parameters of the generators of the Weber Seifert manifold ($L \simeq 1.9927$, $\omega = 3\pi/5$) the expansion converges very slowly. One has to develop the exponential much beyond the rank 300, the calculations are slow and numerical problems begin to appear (it was checked that with smaller values of $L$ and $\omega$, the calculations converge much faster and are correct). Nevertheless, one can try to solve the condition (23), but since we can only use a limited number of terms, we must again use a $\chi^2$ method. therefore there is no real gain with respect to the « direct » method presented in chapter 4. This $\chi^2$ computation gets into numerical troubles very quickly because some of the $a_{mv}{}^{\sigma_1 v_1}$ coefficients may be very large. One way to solve that is to change the normalisation of the cylindrical radial functions. We have not been able to find a way to escape these difficulties.

The failure of the latter attempt is due to the fact that none of the coefficients $a_{mv}{}^{\sigma_1 v_1}$ used in equation (23) are negligible, although this may be due to our normalisation of the cylindrical radial functions. The expansion (D3) can not be limited to a finite number of terms with a good approximation. On the contrary, in the direct method presented in section 5, if we are carefull enough to chose the points into a « small » volume and make sure that their images by the generators are also inside such a volume, the behaviour of the cylindrical functions at low radius, $I(\beta, \mu_z, v, \rho) \approx \rho^v (1 + ...)$, make the basis functions negligible as $|v|$ increases.

We have tried other methods. Among them, one consists to put, inside the fundamental domain, an oscillating charge or a set of oscillating charges as a scalar field source, and look for which frequencies the system becomes resonant. Analytically, using the pretrace formula [6] we get sum rules involving all the eigenstates. More precisely, let $t$ be the time, we consider the space $t \times H^3$, and we put a single unit charge at the point $y$. We define the Green function on $t \times H^3$ as usual by : $(\partial_t^2 - \Delta) G(t, t', x, y) = \delta(t - t') \delta^3(x - y)$ and write it in the form : $G(t, t', x, y) = \delta(t - t' - dist(x, y)) k(x, y)$ where $dist(x, y)$ is the distance in $H^3$ between the points $x$ and $y$. Then, the scalar field in $M$ generated by a point source with charge $\sim f(t)$, is the sum of the fields generated by all the images of the source :

$$\phi(t, x) = \sum_{\gamma \in \Gamma} f(t - dist(x, \gamma y)) k(x, \gamma y)$$

Using the pretrace formula, on can also write : $\phi(t, x) = \sum_s h(t, \lambda_s) \overline{\psi_s}(y) \psi_s(x)$

where $\psi_s(x)$ is the eigenstate of the Laplacian associated to the eigenvalue $\lambda_s = -(1+\beta_s^2)$, and where:
$$h(t,\lambda_s) = \frac{4\pi}{\beta_s} \int_0^\infty f(t-\chi)\, k(0,\chi) \sin(\beta_s \chi)\, sh(\chi)\, d\chi$$

In the case of a rectangular parallelepiped box in Euclidean space, the resonnant states appear clearly. Numerical calculations in this case gives very easily the eigenstates. In the case of the Weber-Seifert manifold, there is no hint of a solution either analytically, or numerically by summing the contributions of all the copies of the fundamental domain.

**Conclusion.**

We have shown that a basis of "cylindrical" functions which are eigenfunctions of the Laplacian and of one generator of the group of deck transformations defining a hyperbolic 3-d compact space, can be used to compute numerically solutions of the Laplacian eigenvalue problem in that space. The method is well suited to compute the lowest eigenmodes. The results concerning the Thurston manifold have been compared to earlier works. The agreement is good but not complete. The method has been applied to compact manifolds having large volume like the Weber-Seifert manifold.

**Ackowledgements.**
I thank J.F. Glicenstein for his comments and for bringing my attention to reference [2] while the first version of this work was finished.

**Appendix A . Elementary transvections and rotations.**

In this appendix we consider infinitesimal transvections and rotations whose base geodesics and axis, respectively, are the coordinate axis.

In cylindrical coordinates, an infinitesimal transvection whose base geodesic is $Oz$ transforms a point of coordinates $(\rho, \varphi, z)$ into a point of coordinates $(\rho, \varphi, z+dz)$ (equations (5)). The corresponding operator acting on a scalar functions of these coordinates is : $T_z = \dfrac{\partial}{\partial z}$.

A rotation of angle $\omega$ around $Oz$ transforms a point of coordinates $(\rho, \varphi, z)$ into a point of coordinates : $(\rho, \varphi + \omega, z)$ . The corresponding operator acting on a scalar function of these coordinates is : $L_z = \dfrac{\partial}{\partial \varphi}$ .

In a transvection along the $Ox$ axis, the distance of a point to this axis is conserved, and its image lies in the « plane » defined by itself and the base geodesic. In a rotation, the distance to the axis is also conserved and the orthogonal projection of this point on the axis is invariant. Taking into account these constraints, the action of an infinitesimal transvection of length $\delta \ll 1$ and of an infinitesimal rotation of angle $\alpha \ll 1$ , is :

$$d\rho = chz\, c_\varphi\, \delta - shz\, s_\varphi\, \alpha$$
$$sh\rho\, d\varphi = -ch\rho\, (chz\, s_\varphi\, \delta + shz\, c_\varphi\, \alpha)$$
$$ch\rho\, dz = sh\rho\, (-shz\, c_\varphi\, \delta + chz\, s_\varphi\, \alpha)$$

where : $c_\varphi = \cos\varphi$ and $s_\varphi = \sin\varphi$ .

from which we deduce dirctly the corresponding operators acting on scalar functions :

$$T_x = chz\, c_\varphi\, \partial_\rho - \frac{ch\rho}{sh\rho} chz\, s_\varphi\, \partial_\varphi - \frac{sh\rho}{ch\rho} shz\, c_\varphi\, \partial_z$$

$$L_x = -shz\, s_\varphi\, \partial_\rho - \frac{ch\rho}{sh\rho} shz\, c_\varphi\, \partial_\varphi + \frac{sh\rho}{ch\rho} chz\, s_\varphi\, \partial_z$$

When the base geodesic is the $Oy$ axis , the same calculations lead to :

$$T_y = chz\, s_\varphi\, \partial_\rho + \frac{ch\rho}{sh\rho} chz\, c_\varphi\, \partial_\varphi - \frac{sh\rho}{ch\rho} shz\, s_\varphi\, \partial_z$$

$$L_y = shz\, c_\varphi\, \partial_\rho - \frac{ch\rho}{sh\rho} shz\, s_\varphi\, \partial_\varphi - \frac{sh\rho}{ch\rho} chz\, c_\varphi\, \partial_z$$

These operators can be written in spherical coordinates. For that purpose we use the geometrical relations of rectangular triangles : $ch\chi = ch\rho\, chz$ , $thz = c\, th\chi$ and $sh\rho = s\, sh\chi$ which link the two sets of coordinates, where $c = \cos\theta$ and $s = \sin\theta$ . We obtain :

$$T_z = c\, \partial_\chi - \frac{ch\chi}{sh\chi} s\, \partial_\theta \quad , \quad L_z = \frac{\partial}{\partial \varphi}$$

$$T_x = s\, c_\varphi\, \partial_\chi + \frac{c\, c_\varphi}{th\chi}\, \partial_\theta - \frac{s_\varphi}{s\, th\chi}\, \partial_\varphi$$

$$L_x = -s_\varphi\, \partial_\theta - \frac{c}{s}\, c_\varphi\, \partial_\varphi$$

One directly checks the following commutation relations, as expected from the commutation relations of the Lie Algebra of $SO(3,1)$ :

$$[T_x, T_y] = L_z \quad , \quad [T_y, T_z] = L_x \quad , \quad [T_z, T_x] = L_y$$
$$[T_x, L_y] = -T_z \quad , \quad [L_x, T_y] = -T_z \quad , \quad [T_z, L_x] = -T_y$$
$$[T_z, L_y] = T_x \quad , \quad [L_z, T_x] = -T_y \quad , \quad [L_z, T_y] = T_x$$
$$[L_x, L_y] = -L_z \quad , \quad [L_y, L_z] = -L_x \quad , \quad [L_z, L_x] = -L_y$$

At last we have the relation (Casimir Operator) :
$$(T_x^2 + T_y^2 + T_z^2) - (L_x^2 + L_y^2 + L_z^2) = \Delta$$

The action, on a scalar function, of a motion which is the product of a finite transvection of length $l$ and base geodesic $Oz$, and a finite rotation of axis $Oz$ and angle $\theta$ has the form : $e^{lT_z + \theta L_z}$ . If the axis of the motion is $Ox$, the operator is of course : $e^{lT_x + \theta L_x}$ .

**Appendix B , Recurrence relations for the radial cylindrical Functions.**

In cylindrical coordinates, we look for solutions of the differential equation : $\Delta \varphi_{\sigma v} = -(1 + \beta^2) \varphi_{\sigma v}$ , of the form : $\varphi_{\sigma,v} = I_{\sigma,v}(\rho)\, e^{\sigma z + i v \varphi}$ , where $v$ is an integer and $\sigma$ is a complex number.

The radial cylindrical functions $I_{\sigma,v}(\rho)$ are solutions of the differential equation :

$$\frac{1}{sh\rho\, ch\rho}\, \partial_\rho (sh\rho\, ch\rho\, \partial_\rho I_{\sigma,v}) + \frac{\sigma^2}{ch^2\rho} I_{\sigma,v} - \frac{v^2}{sh^2\rho} I_{\sigma,v} = -(1+\beta^2) I_{\sigma,v}$$

Setting : $I' = \partial_\rho I_{\sigma,v}$ and $I'' = \partial_{\rho\rho} I_{\sigma,v}$ , the equation is :

$$I'' + \left( \frac{ch\rho}{sh\rho} + \frac{sh\rho}{ch\rho} \right) I' + \frac{\sigma^2}{ch^2\rho} I - \frac{v^2}{sh^2\rho} I = -(1+\beta^2) I$$

We derive this equation and with : $\psi = \partial_\rho I_{\sigma,v} = I'$ , one gets :

$$\psi'' + \left( \frac{ch\rho}{sh\rho} + \frac{sh\rho}{ch\rho} \right) \psi' + \frac{\sigma^2}{ch^2\rho} \psi - \frac{v^2}{sh^2\rho} \psi + \left( \frac{1}{ch^2\rho} - \frac{1}{sh^2\rho} \right) \psi - 2\sigma^2 \frac{sh\rho}{ch^3\rho} I + 2v^2 \frac{ch\rho}{sh^3\rho} I = -(1+\beta^2) \psi$$

The expansion (15) suggests to try a solution of the form : $\psi = \phi + f\, I_{\sigma,v}$ , where : $f = a\, \dfrac{ch\rho}{sh\rho} + b\, \dfrac{sh\rho}{ch\rho}$ . Replacing into the differential equation gives :

$$\left[\phi''+\left(\frac{ch\rho}{sh\rho}+\frac{sh\rho}{ch\rho}\right)\phi'\right]+\left[\frac{1}{ch^2\rho}-\frac{1}{sh^2\rho}+\frac{\sigma^2}{ch^2\rho}-\frac{v^2}{sh^2\rho}+\frac{2b}{ch^2\rho}-\frac{2a}{sh^2\rho}\right]\phi$$

$$+\left[2(b^2-\sigma^2)\frac{sh\rho}{ch^3\rho}+2(v^2-a^2)\frac{ch\rho}{sh^3\rho}\right]I=-(1+\beta^2)\phi$$

This equation simplifies if we choose :
$$b=\pm\sigma \quad , \quad a=\pm v$$
and reduces to :
$$\phi''+\left(\frac{ch\rho}{sh\rho}+\frac{sh\rho}{ch\rho}\right)\phi'+\frac{(b+1)^2}{ch^2\rho}\phi-\frac{(a+1)^2}{sh^2\rho}\phi=-(1+\beta^2)\phi$$

which is the defining differential equation of the radial functions.
There are two possibilities. We define $\overline{v}=|v|$ , then :

$$a=+\overline{v} \quad : \quad \partial_\rho I_{\sigma,v}=(\overline{v}\frac{ch\rho}{sh\rho}+b\frac{sh\rho}{ch\rho})I_{\sigma,v}+c\,I_{b+1,\overline{v}+1}$$

$$a=-\overline{v} \quad : \quad \partial_\rho I_{\sigma,v}=(-\overline{v}\frac{ch\rho}{sh\rho}+b'\frac{sh\rho}{ch\rho})I_{\sigma,v}+d\,I_{b'+1,\overline{v}-1}$$

where $c$ and $d$ are constants, and : $b, b'=\pm\sigma$. When $v=0$ only the first relation remains.
The coefficients $c$ and $d$ are determined by using the expansion (15). This can be done at the lowest order, but has also been checked at the next order. The result is :

$$c=\frac{(1-\beta^2-\sigma^2-\overline{v}^2)}{2(\overline{v}+1)}-(1+b) \quad , \quad d=2\overline{v}$$

Of course the value of these coeficients depends on the normalisation of the radial functions implied by the choice (15).
These relations look like the Bessel function recurrence relations.

### Appendix C. Expressions of the elementary motions applied to the set of basis functions.

A screw motion in $H^3$ is the product of a transvection of length $l$ and a rotation of angle $\theta$ around an axis which is the base geodesic of the transvection, the order in which these two operations are performed does not matter because they commute. We define elementary screw motions with respect to the coordinate axis $Oz$ and $Ox$ respectively by $V_z=l\,T_z+\theta\,L_z$ , $(l,\theta\ll 1)$ , and $V_x=l\,T_x+\theta\,L_x$ where the operators $T_z$ , $L_z$ , $T_x$ , $L_x$ have been calculated in Appendix A. Their actions on the basis functions have simple expressions that are given in the present Appendix.

Spherical coordinates.

In spherical coordinates the functions $\varphi_{lm}=\phi_\beta^l\,P_l^m\,e^{im\varphi}$ , where $\phi_\beta^l$ has been defined in section 3 and $P_l^m$ are the Legendre polynomials, satisfy the equation $\Delta\varphi_{lm}=-(1+\beta^2)\varphi_{lm}$ .

Using the recurrence relations of the Legrendre functions [5], the functions $\phi_\beta^l$ are related by:

$$\frac{ch\chi}{sh\chi} \phi_\beta^l = \frac{1}{2l+1} \left[ \phi_\beta^{l-1} + \left(\beta^2 + (l+1)^2\right) \phi_\beta^{l+1} \right] \tag{C1}$$

$$\partial_\chi \phi_\beta^l = \frac{1}{2l+1} \left[ l\, \phi_\beta^{l-1} - (l+1)\left(\beta^2 + (l+1)^2\right) \phi_\beta^{l+1} \right] \tag{C2}$$

with these equations, and the recurrence relations of the Legendre polynomials, one obtains the following simple relations:

$$(2l+1)\, T_z(\phi_\beta^l P_l^m) = (l+m)\, \phi_\beta^{l-1} P_{l-1}^m - (l+1-m)\,(\beta^2+(l+1)^2)\, \phi_\beta^{l+1} P_{l+1}^m \tag{C3a}$$

$$L_z(\phi_\beta^l P_l^m e^{im\varphi}) = i\nu\, \phi_\beta^l P_l^m e^{im\varphi} \tag{C3b}$$

$$2(2l+1)\, T_x(\phi_\beta^l P_l^m e^{im\varphi}) = \phi_\beta^{l-1}\left[ P_{l-1}^{m+1} e^{i(m+1)\varphi} - (l+m)(l+m-1) P_{l-1}^{m-1} e^{i(m-1)\varphi} \right]$$
$$+ (\beta^2+(l+1)^2)\, \phi_\beta^{l+1}\left[ P_{l+1}^{m+1} e^{i(m+1)\varphi} - (l+1-m)(l+2-m) P_{l+1}^{m-1} e^{i(m-1)\varphi} \right]$$
$$\tag{C4a}$$

$$L_x(\phi_\beta^l P_l^m e^{im\varphi}) = \frac{i}{2} \phi_\beta^l P_l^{m+1} e^{i(m+1)\varphi} + \frac{i}{2}(l+m)(l+1-m)\, \phi_\beta^l P_l^{m-1} e^{i(m-1)\varphi} \tag{C4b}$$

$L_z$ and $L_x$ conserve the total angular momentum $l$, as expected.

Cylindrical coordinates.

We define: $\varphi_{\sigma,\nu} = I_{\sigma,\nu}(\rho)\, e^{\sigma z + i\nu\varphi}$ where $I_{\sigma,\nu}(\rho)$ is the radial function defined in Appendix B. By construction these functions satisfy $\Delta\varphi_{\sigma\nu} = -(1+\beta^2)\varphi_{\sigma\nu}$.

We note $c_+$ and $c_-$ the coefficients of the Appendix B recurrence relations when $b = +\sigma$ and $b = -\sigma$ respectively.

When $\nu > 0$, one obtains:
$$4V_x\, \varphi_{\sigma,\nu} = (l+i\theta) c_+ \varphi_{\sigma+1,\nu+1} + (l-i\theta) c_- \varphi_{\sigma-1,\nu+1} + (l-i\theta) d\, \varphi_{\sigma+1,\nu-1} + (l+i\theta) d\, \varphi_{\sigma-1,\nu-1} \tag{C5}$$

and when $\nu < 0$, with $\bar{\nu} = |\nu|$:
$$4V_x\, \varphi_{\sigma,\nu} = (l-i\theta) c_+ \varphi_{\sigma+1,-(\bar{\nu}+1)} + (l+i\theta) c_- \varphi_{\sigma-1,-(\bar{\nu}+1)} + (l+i\theta) d\, \varphi_{\sigma+1,-(\bar{\nu}-1)} + (l-i\theta) d\, \varphi_{\sigma-1,-(\bar{\nu}-1)}$$
$$\tag{C6}$$

Which is the same formula as in the case $\nu > 0$ with the sign of $\theta$ reverse.

When $\nu = 0$ the expression is slightly different:
$$4V_x\, \varphi_{\sigma,\nu=0} = (l+i\theta) c_+ I_{\sigma+1,1}\, e^{(\sigma+1)z+i\varphi} + (l-i\theta) c_-\, I_{\sigma-1,1}\, e^{(\sigma-1)z+i\varphi}$$
$$+ (l-i\theta) c_+ I_{\sigma+1,1}\, e^{(\sigma+1)z-i\varphi} + (l+i\theta) c_-\, I_{\sigma-1,1}\, e^{(\sigma-1)z-i\varphi} \tag{C7}$$

The action of $V_z$ is very simple:
$$V_z\, \varphi_{\sigma,\nu} = (l\partial_z + \theta\partial_\varphi)\, \varphi_{\sigma,\nu} = (\sigma l + i\nu\theta)\, \varphi_{\sigma,\nu}$$

The full action of this motion is:
$$e^{V_z}\, \varphi_{\sigma,\nu} = I_{\sigma,1}\, e^{\sigma(z+l)+i\nu(\varphi+\theta)} = e^{l\sigma+i\nu\theta}\, \varphi_{\sigma,\nu} \tag{C8}$$

which shows that the functions $\varphi_{\sigma,\nu}$ are invariant by $\gamma_0$ only if: $L\sigma + i\nu\omega = 2i\pi m$.

## $\gamma_0$ invariant functions in spherical coordinates.

Recall that $\gamma_0$ has been defined as the generator of the group of deck transformations whose base geodesic defines the $Oz$ axis. We make the following expansion :

$$I(\mu_z, \nu, \rho) \exp(i(\mu_z z + \nu\varphi)) = \sum_{l,m} a_{lm} \phi_\beta^l(\chi) P_l^m(\theta) \exp(im\varphi)$$

Which, with $m = \nu$ by orthogonality of the functions $e^{i\nu\varphi}$ on $[0, 2\pi]$, reduces to ($\bar{\nu} = |\nu|$) :

$$I(\mu_z, \nu, \rho) \exp(i\mu_z z) = \sum_{l \geq \bar{\nu}} a_l \phi_\beta^l(\chi) P_l^\nu(\theta)$$

The $a_l$ coefficients are calculated by imposing :

$$T_z^2 I(\mu_z, \nu, \rho) \exp(i\mu_z z) = \sum_{l \geq \bar{\nu}} a_l T_z^2 \phi_\beta^l P_l^\nu = -\mu_z^2 \sum_{l \geq \bar{\nu}} a_l \phi_\beta^l P_l^\nu$$

and where, from the above expression for $T_z$, we have :

$$(2l+1) T_z^2 (\phi_\beta^l P_l^\nu) = \frac{(l+\nu)(l+\nu-1)}{(2l-1)} \phi_\beta^{l-2} P_{l-2}^\nu$$

$$- \left[ \frac{(l+\nu)(l-\nu)}{(2l-1)} (\beta^2 + l^2) + \frac{(l+\nu+1)(l+1-\nu)}{(2l+3)} (\beta^2 + (l+1)^2) \right] \phi_\beta^l P_l^\nu$$

$$+ \frac{(l+1-\nu)(l+2-\nu)}{(2l+3)} (\beta^2 + (l+1)^2)(\beta^2 + (l+2)^2) \phi_\beta^{l+2} P_{l+2}^\nu$$

which gives a recurrence relations between the coefficients $a_l$, $a_{l-2}$ and $a_{l+2}$.
Using low order expansions of the $\phi_\beta^l$ and the expressions of $P_l^l$ and $P_l^{l-1}$, the first coefficients are related by : $\quad \dfrac{a_{\nu+1}}{a_\nu} = i\mu \dfrac{2\nu+3}{2\nu+1}$ , if $\nu \geq 0$ , and $\quad \dfrac{a_{\bar{\nu}+1}}{a_{\bar{\nu}}} = i\mu(2\bar{\nu}+3)$ , if $\nu < 0$ .

## Complement on the radial functions of the spherical coordinates.

The relation (C3a) shows that $T_z$ projects a state $\varphi_{lm} = \phi_\beta^l P_l^m e^{im\varphi}$ on the states $\varphi_{l\pm1, m}$ by conserving $m$. If the points lie on the $Oz$ axis, we have : $\cos(\theta) = 1$ and therefore : $P_l^0 = 1$ and $P_l^m = 0$ if $m \neq 0$ whatever the value of $l$. Then one can write :

$$e^{yT_z} \phi_\beta^l(x) = \phi_\beta^l(x+y) = \sum_{l_1 \geq 0} r_{l_1}^l(y) \phi_\beta^{l_1}(x)$$

where the coefficients $r_{l_1}^l(y)$ have to be determined.
By symetry the role of $x$ and $y$ can be reversed, and one can write :

$$\phi_\beta^l(x+y) = \sum_{l_1, l_2 \geq 0} r_{l_1 l_2}^l \phi_\beta^{l_1}(x) \phi_\beta^{l_2}(y)$$

By reversing again the role of $x$ and $y$ one has : $r_{l_1 l_2}^l = r_{l_2 l_1}^l$ .
If $y = 0$, $\phi_\beta^{l_2}(0) = 0$ if $l_2 > 0$ and $\phi_\beta^0(0) = 1$, therefore : $r_{l_1 0}^l = \delta_{l_1}^l$ .

The other coefficients $r_{l_1 l_2}^l$ can be obtained by successive derivations. For instance if one derivates once the above expansion with respect to $y$ at the point $y = 0$, and uses the relations (C1) and (C2), one obtains : $r_{l-1,1}^l = 3l/(2l+1)$ , $r_{l+1,1}^l = -\dfrac{3(l+1)}{(2l+1)}(\beta^2 + (l+1)^2)$ , otherwise : $r_{l_1,1}^l = 0$ .

**Appendix D. The case of the Weber-Seifert manifold and its high degree of symetry.**

The base geodesics of all the generators of the group of deck transformations of the Weber-Seifert manifold go through the origin.
Let us assume that the base geodesic of the first generator $\gamma_1$ lies in the $xOz$ plane. It makes an angle $\alpha$ with the $Oz$ axis, which is according to our definition, the base geodesic of $\gamma_0$, such that $\cos(\alpha) = 1/\sqrt{5}$.
The motions corresponding to the generators are all the same up to a rotation, and we can write :
$$\gamma_1 = e^{-\alpha L_y} \gamma_0 e^{\alpha L_y} = e^{-\alpha L_y} e^{lT_z + \theta L_z} e^{\alpha L_y} \tag{D1}$$
The other generators can be obtained from $\gamma_1$ by a rotation of $2\pi/5$ around the $Oz$ axis.
This equation can be transformed using Hadamard's formula. With $V_z = l T_z + \theta L_z$, and $V_x = l T_x + \theta L_x$ one has :
$$\gamma_1 = \exp\left(V_z - \alpha[L_y, V_z] + \tfrac{1}{2}\alpha^2 [L_y,[L_y,V_z]] - \tfrac{1}{3!}\alpha^3 [L_y,[L_y,[L_y,V_z]]] + ...\right)$$
Using the following commutation relations deduced from those of appendix A
$$[L_y, V_z] = -V_x \quad \text{and} \quad [L_y, V_x] = V_z$$
the generator $\gamma_1$ can be simply rewritten :
$$\gamma_1 = e^{c_\alpha V_z + s_\alpha V_x} \tag{D2}$$
where : $c_\alpha = \cos\alpha$ , $s_\alpha = \sin\alpha$ , or equivalently : $\gamma_1 = e^{c_\alpha T_z + s_\alpha T_x} e^{c_\alpha L_z + s_\alpha L_x}$ .

We call $\varphi_{\mu\nu}$ the cylindrical functions invariant by $\gamma_0$ satisfying the equation $\Delta \varphi_{\mu\nu} = -(1+\beta^2)\varphi_{\mu\nu}$ and of the form : $\varphi_{\mu,\nu} = I_{\mu,\nu}(\rho) e^{i\mu z + i\nu\varphi}$ where $\mu L + \nu \omega = 2\pi m$ .
According to the above expression for $\gamma_1$ and the formulae of Appendix C, we have the following expansion :
$$\varphi_{\mu\nu}(\gamma_1 x) = a_{\mu\nu}^{\sigma_1 \nu_1} \varphi_{\sigma_1 \nu_1}(x) \tag{D3}$$
where : $\sigma_1 = i\mu + p$ with $p \in \mathbb{Z}$ .
From the defining equations we have : $\overline{\varphi_{\sigma\nu}} = \varphi_{\bar\sigma, -\nu}$ from which we deduce that :
$a_{\bar\sigma, -\nu}^{\bar\sigma_1, -\nu_1} = \overline{a_{\sigma\nu}^{\sigma_1 \nu_1}}$ .

## Appendix E.

The following figure represents an icosahedron as if it were a stereographic projection but without respecting the lengths. Faces are named with letters and summits with numbers.

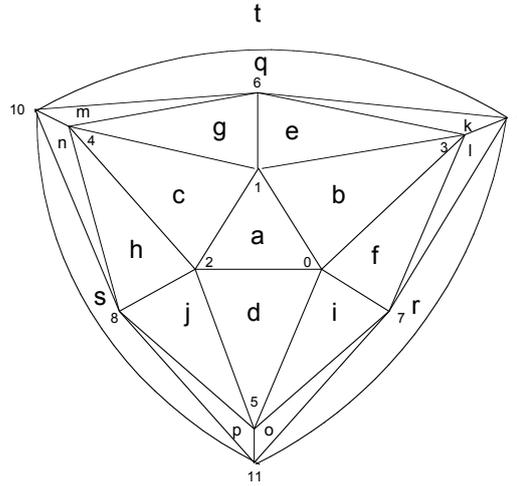

Each of the following tables shows how the faces of the icosahedron are associated to build a manifold. Each line gives the corresponding faces, and how the summits are associated. The first 3 summits belong to the face in the first column, and the next 3 to the face in the second column. For instance in the next table, the correspondance between the faces $a$ and $d$ is done according to : $0 \to 5$, $1 \to 2$, $2 \to 0$. The next 6 tables correspond to the solutions 14, 12, 13, 10, 9, 11 of [7].

| faces | | summits | | | summits | | |
|---|---|---|---|---|---|---|---|
| a | d | 0 | 1 | 2 | 5 | 2 | 0 |
| b | h | 0 | 1 | 3 | 4 | 8 | 2 |
| c | i | 4 | 1 | 2 | 7 | 0 | 5 |
| e | s | 6 | 1 | 3 | 11 | 10 | 8 |
| f | o | 0 | 7 | 3 | 7 | 11 | 5 |
| g | r | 4 | 1 | 6 | 11 | 7 | 9 |
| j | n | 8 | 5 | 2 | 10 | 4 | 8 |
| k | m | 6 | 9 | 3 | 10 | 4 | 6 |
| l | t | 9 | 7 | 3 | 11 | 10 | 9 |
| p | q | 8 | 5 | 11 | 6 | 9 | 10 |

| faces | | summits | | | | summits | | |
|---|---|---|---|---|---|---|---|---|
| a | d | 0 | 1 | 2 | | 2 | 0 | 5 |
| b | j | 0 | 1 | 3 | | 5 | 2 | 8 |
| c | k | 4 | 1 | 2 | | 9 | 6 | 3 |
| e | i | 6 | 1 | 3 | | 0 | 7 | 5 |
| f | t | 0 | 7 | 3 | | 10 | 11 | 9 |
| g | s | 4 | 1 | 6 | | 8 | 11 | 10 |
| h | o | 4 | 8 | 2 | | 11 | 5 | 7 |
| l | r | 9 | 7 | 3 | | 11 | 9 | 7 |
| m | n | 4 | 10 | 6 | | 10 | 8 | 4 |
| p | q | 8 | 5 | 11 | | 9 | 10 | 6 |

| faces | | summits | | | | summits | | |
|---|---|---|---|---|---|---|---|---|
| a | d | 0 | 1 | 2 | | 2 | 0 | 5 |
| b | j | 0 | 1 | 3 | | 5 | 2 | 8 |
| c | l | 4 | 1 | 2 | | 3 | 7 | 9 |
| e | n | 6 | 1 | 3 | | 4 | 8 | 10 |
| f | q | 0 | 7 | 3 | | 6 | 10 | 9 |
| g | t | 4 | 1 | 6 | | 9 | 10 | 11 |
| h | s | 4 | 8 | 2 | | 11 | 10 | 8 |
| i | r | 0 | 5 | 7 | | 7 | 9 | 11 |
| k | p | 6 | 9 | 3 | | 5 | 8 | 11 |
| m | o | 4 | 10 | 6 | | 5 | 11 | 7 |

| faces | | summits | | | | summits | | |
|---|---|---|---|---|---|---|---|---|
| a | d | 0 | 1 | 2 | | 2 | 0 | 5 |
| b | j | 0 | 1 | 3 | | 5 | 2 | 8 |
| c | q | 4 | 1 | 2 | | 10 | 6 | 9 |
| e | r | 6 | 1 | 3 | | 11 | 7 | 9 |
| f | t | 0 | 7 | 3 | | 11 | 9 | 10 |
| g | m | 4 | 1 | 6 | | 6 | 4 | 10 |
| h | l | 4 | 8 | 2 | | 3 | 9 | 7 |
| i | k | 0 | 5 | 7 | | 6 | 9 | 3 |
| n | o | 4 | 8 | 10 | | 7 | 5 | 11 |
| p | s | 8 | 5 | 11 | | 10 | 11 | 8 |

| faces | | summits | | | | summits | | |
|---|---|---|---|---|---|---|---|---|
| a | j | 0 | 1 | 2 | | 2 | 5 | 8 |
| b | i | 0 | 1 | 3 | | 5 | 7 | 0 |
| c | n | 4 | 1 | 2 | | 10 | 8 | 4 |
| d | o | 0 | 5 | 2 | | 11 | 7 | 5 |
| e | l | 6 | 1 | 3 | | 3 | 9 | 7 |
| f | r | 0 | 7 | 3 | | 7 | 9 | 11 |
| g | q | 4 | 1 | 6 | | 6 | 10 | 9 |
| h | p | 4 | 8 | 2 | | 8 | 5 | 11 |
| k | t | 6 | 9 | 3 | | 11 | 10 | 9 |
| m | s | 4 | 10 | 6 | | 11 | 8 | 10 |

| faces | | summits | | | summits | | |
|---|---|---|---|---|---|---|---|
| a | j | 0 | 1 | 2 | 2 | 5 | 8 |
| b | l | 0 | 1 | 3 | 9 | 3 | 7 |
| c | o | 4 | 1 | 2 | 5 | 11 | 7 |
| d | k | 0 | 5 | 2 | 6 | 3 | 9 |
| e | n | 6 | 1 | 3 | 4 | 8 | 10 |
| f | s | 0 | 7 | 3 | 11 | 10 | 8 |
| g | t | 4 | 1 | 6 | 11 | 9 | 10 |
| h | q | 4 | 8 | 2 | 10 | 9 | 6 |
| i | m | 0 | 5 | 7 | 4 | 10 | 6 |
| p | r | 8 | 5 | 11 | 7 | 11 | 9 |

| faces | | summits | | | summits | | |
|---|---|---|---|---|---|---|---|
| a | j | 0 | 1 | 2 | 2 | 5 | 8 |
| b | q | 0 | 1 | 3 | 9 | 10 | 6 |
| c | k | 4 | 1 | 2 | 9 | 6 | 3 |
| d | t | 0 | 5 | 2 | 11 | 10 | 9 |
| e | p | 6 | 1 | 3 | 5 | 11 | 8 |
| f | g | 0 | 7 | 3 | 4 | 6 | 1 |
| h | r | 4 | 8 | 2 | 7 | 9 | 11 |
| i | n | 0 | 5 | 7 | 10 | 4 | 8 |
| l | o | 9 | 7 | 3 | 7 | 5 | 11 |
| m | s | 4 | 10 | 6 | 10 | 11 | 8 |

The Weber Seifert is described by the next table. The associated faces are diametrically opposed.

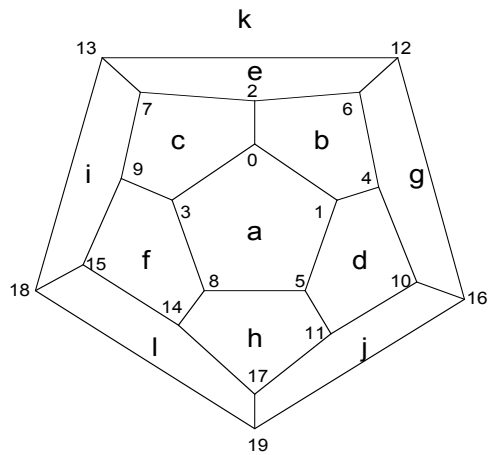

| faces | | summits | | | | | summits | | | | |
|---|---|---|---|---|---|---|---|---|---|---|---|
| a | k | 0 | 1 | 5 | 8 | 3 | 16 | 19 | 18 | 13 | 12 |
| b | l | 0 | 1 | 4 | 6 | 2 | 18 | 15 | 14 | 17 | 19 |
| c | j | 0 | 3 | 9 | 7 | 2 | 17 | 19 | 16 | 10 | 11 |
| d | i | 1 | 4 | 10 | 11 | 5 | 13 | 18 | 15 | 9 | 7 |
| e | h | 6 | 12 | 13 | 7 | 2 | 8 | 5 | 11 | 17 | 14 |
| f | g | 3 | 8 | 14 | 15 | 9 | 10 | 16 | 12 | 6 | 4 |